\documentstyle[preprint,eqsecnum,aps]{revtex}
\tightenlines

\input{psfig.sty}

\begin{document}

\draft
\title{Polarization Control of the Non-linear Emission on Semiconductor Microcavities}

\author{M.D. Mart\'in$^*$, G. Aichmayr, and L. Vi\~na}
\address{
Departamento de F\'isica de Materiales C-IV, Universidad Aut\'onoma de Madrid,
Cantoblanco,
E-28049 Madrid, Spain}

\author{R. Andr\'e}
\address{Lab. Spectrometrie Physique (CNRS), Univ. Joseph Fourier 1, F-38402 Grenoble, France}
\maketitle
\begin{abstract}

The degree of circular polarization ($\wp$) of the non-linear emission in 
semiconductor microcavities is controlled by changing the 
exciton-cavity detuning. The polariton relaxation towards \textbf{K} $\sim 0$ 
cavity-like states is governed by final-state stimulated scattering. 
The helicity of the emission is selected due to the lifting of the degeneracy of the
$\pm 1$ spin levels at \textbf{K} $\sim 0$. 
At short times after a pulsed excitation $\wp$ reaches very large values, either positive or negative, 
as a result of stimulated scattering to the spin level of lowest energy ($+1/-1$ spin for positive/negative detuning).

\end{abstract}


Semiconductor microcavities have attracted increasing interest in the last decade because
they allow a precise control of the radiation-matter interaction.
This interaction is strongest when the characteristic frequencies of photons (radiation)
and excitons (matter) are brought into resonance. Two different regimes can be
established under this resonance condition: 
the strong and weak coupling regimes. The largest effort has been devoted 
to the study of the strong coupling regime (SCR), 
in which the eigenstates of the system are no longer pure exciton or photon but 
a superposition of both, known as cavity-polaritons.\cite{Weisbuch92,Houdre95} 
The resonant frequencies of excitons and photons are split, 
leading to the so-called Rabi splitting, in analogy with atomic cavities.\cite{Haroche89} 
Only in the last years it has been possible to 
observe the polariton non-linear emission in both, 
III-V\cite{Senellart99,Stevenson00,Dasbach00} and 
II-VI\cite{Dang98,Bleuse98,Boeuf00,Mueller00} semiconductor microcavities. 

Other issue that has drawn a lot of attention in the non-linear SCR is 
the existence of a polariton-polariton scattering mechanism stimulated 
by the final state population. This mechanism will be active in a bosonic system, 
such as cavity polaritons, as soon as the final state population approaches unity. 
Clear experimental evidences of this stimulated scattering have been reported 
recently in the literature.\cite{Stevenson00,Savvidis00} In those experiments, 
the parametric scattering was enhanced by a convenient choice of the angle of 
incidence of the excitation beams. The result is a macroscopic polariton 
occupancy ("condensation") of the states at \textbf{K} $\sim 0$ 
and \textbf{K} $\sim 2 \textbf{k}_{pump} $, 
where $\textbf{k}_{pump}$ is the incident pump wave vector.

Concomitantly, a rekindled interest on the carriers' spin in semiconductor structures 
has given rise to a new field, spintronics, which explores the possibility of 
designing new spin-based devices, useful for advanced applications such as
optical memories and switches, quantum cryptography and quantum computing.
The degree of circular polarization of the emission is directly 
related with the spin of the elementary excitations of the system, defined as
the third component of the total angular momentum. The spin 
relaxation processes of excitons, electrons and holes have been extensively 
studied in the last decades.\cite{Damen91,Vinattieri93,Ferreira94,Maialle93}
In the particular case of 
cavity polaritons, due to the mixed photon-exciton character, significant 
changes on their spin dynamics with respect to bare quantum wells are 
expected. Nevertheless, the spin has been considered only 
very recently.\cite{Savvidis00,Tartakovskii99,Renucci01,Martin01}

In this paper, the fundamental issue of polariton spin dynamics is investigated
under non-resonant excitation conditions, which resemble those expected in
real devices.  We demonstrate that the polariton spin plays a crucial role 
in the stimulated scattering process, which could lead to an 
exciton-polariton condensate and polariton lasing. We establish that
semiconductor microcavities offer unique possibilities to control the helicity
of the light emission, what could be exploited to develop ultrafast optical
polarization switches. This control, determined by the exciton-cavity detuning, results
from a breaking of the degeneracy of the spin-up and -down cavity-like states: 
the ground state of the system is spin-up (-down) at positive (negative) detunings.

The sample under study is a $\lambda/2$ Cd$_{0.40}$Mg$_{0.60}$Te. 
A slight wedge in the cavity thickness, obtained by suppression of sample
rotation during molecular-beam-epitaxy growth, allows tuning of the cavity 
and the exciton by moving the excitation spot across the wafer.
In the antinode position of the 
electromagnetic standing wave 2 CdTe QWs of 90 {\AA} are placed. 
The top/bottom cavity mirrors are Distributed Bragg Reflectors (DBRs) made of 
17.5/23 pairs of alternating layers of Cd$_{0.40}$Mg$_{0.60}$Te 
and Cd$_{0.75}$Mg$_{0.25}$Te. The cavity finesse, extracted 
from cw-reflectivity measurements, amounts to $\sim 1200$, 
assuring the excellent quality of the sample. The Rabi splitting 
characteristic of the microcavity is also determined by cw-reflectivity 
measurements and amounts to $\sim 10.5$ meV.

The experiments are made in back-scattering geometry with pulses provided by a Ti:Sapphire laser,
and the sample kept at 5 K.
The photoluminescence (PL) is collected, for  $K \leq 1 {\times} 10 ^{4} cm ^{-1}$, and 
time-resolved using an up-conversion spectrometer with 2 ps resolution. 
For polarization-resolved 
measurements, $\lambda/4$ plates are used to analyze the emitted PL into its $\sigma ^+$- and $\sigma ^-$-polarized 
components, after excitation with $\sigma ^+$ light pulses. 
The circular degree of polarization of the PL is defined as
$\wp=\frac{I^+-I^-}{I^++I^-}$, 
where $I^{+/-}$ is the intensity of the $\sigma ^{+/-}$ 
component of the PL. $\wp$ will be denoted in the following as polarization.
The non-resonant 
excitation energy is tuned to the first reflectivity dip above the stop-band 
of the DBRs, $\sim 90$ meV higher than the emission energy of the cavity-like 
polariton branch, to assure the same excitation conditions in all the experiments. 
The created excitons relax very fast ($\sim 100 fs$) by optic phonon emission to
polariton states. The scattering processes to thermalize the polaritons close to
\textbf{K} $\sim 0$ are profusely discussed in Refs. \cite{Dang98} and \cite {Savvidis02}, where
exciton-exciton Coulomb scattering is shown to play a key role.

We have studied the time evolution of the polariton PL as a function of the cavity-exciton 
detuning ($\delta = E_{C} - E_{X}$) and of the average excitation power density. 
Figure 1 shows a contour plot that compiles the PL spectra at 10 ps, 
measured at different points of the sample. The PL shows a clearly resolved doublet, 
which allows to determine the energy splitting between the two normal modes. 
A clear anticrossing between the bare states is observed, leading to a Rabi splitting 
of $\sim 9.5$ meV, comparable to that observed in cw experiments. 
We will concentrate on the results at $\delta = \pm 10$ meV, which show the 
highest values of $\wp$. At intermediate detunings the dynamics 
of the light emission evolves monotonously between these two cases; 
however, $\wp$ presents a rich evolution, with striking 
phenomena such as the existence of a time-plateau, 
which will be presented elsewhere.\cite{Aichmayrxx}

The integrated intensity of the photon-like branch emission in a point of the sample
characterized by $\delta \sim -10$ meV is depicted in the 
inset of Fig. 1, for a delay time of $\sim 20$ ps. An exponential rise of 
the emission intensity with increasing excitation density is observed 
above 7.5 $W/cm^{2}$. A similar exponential 
growth has been reported and interpreted in 
terms of final-state stimulated scattering, as is characteristic of a 
bosonic system when the final-state (\textbf{K} $\sim 0$) occupancy 
approaches unity.\cite{Baumberg00} 
Our results provide an experimental evidence of a very efficient 
polariton-polariton stimulated scattering, even under non-resonant excitation. 
In the case of positive detuning the excitation power dependence is more complicated. 
For small excitation densities the emission arises mainly from 
the lower polariton branch (LPB) states. 
However, a crossover is observed around $7 W/cm^{2}$, when the emission of 
the upper polariton branch (UPB) exceeds that of the LPB. A similar exponential growth of the 
integrated intensity versus excitation density is then observed, 
indicating the existence of a stimulation process.

In the following we will consider only the non-linear emission regime and 
concentrate on the spin dynamics of cavity polaritons for both, positive 
and negative detunings. The polarization-resolved time evolution of the 
emission from the cavity-like polariton states for an excitation density 
of $18 W/cm^{2}$ is shown in Fig. 2 (a/b) for a detuning of +10/-10 meV.
For $\delta > 0$ (Fig. 2a), the emission intensity of the 
$\sigma ^+$ component of the PL (solid circles) is much bigger than 
that of the $\sigma ^-$-polarized one (open circles). 
This intensity difference gives rise to a non-vanishing polarization of the emission, 
whose time evolution is depicted in Fig. 2 (c). The initial polarization is $\sim 30 $ \% 
and it increases up to $\sim 90 \%$ at $\sim 15 $ ps. This rise of $\wp$, 
which has been also reported recently for III-V 
microcavities,\cite{Renucci01,Martin01} is markedly different to the monotonous decrease 
observed for bare excitons. It can be qualitatively understood considering 
that the relaxation to \textbf{K} $\sim 0$ states, driven by the polariton-polariton 
stimulated scattering, is spin selective and occurs in a much shorter time 
scale than any spin relaxation. This will imply that the slightly 
unbalanced +1 spin population, created at $\textbf{K} > 0$ by a $\sigma ^+$-polarized 
pulse, is mirrored at \textbf{K} $\sim 0$, creating a seed for stimulation 
only of the +1 spin population. However, this argument alone is not sufficient 
to explain the results for $\delta < 0$ as we will show below. After reaching 
its maximum value, $\wp$ decreases to zero, as a result of 
two different processes. The first one is the rapid $\sigma ^+$-polarized stimulated emission, 
which results in a considerable reduction of the +1 spin population. 
The second process, slower than the first one, is the conventional spin relaxation, 
which tends to equalize both spin populations.

The behavior is quite different for $\delta < 0$. 
The polarization-resolved time evolution of the cavity-like mode emission (Fig. 2 b) 
shows that the intensity of the $\sigma ^-$ emission (open circles) 
is larger than that of the $\sigma ^+$-polarized one (solid circles). 
Therefore, the emission is counter-polarized 
with the excitation. This is made evident in the time evolution of $\wp$, 
depicted in Fig. 2 (d). The initial value of $\wp$ is $\sim 50 \%$ but it changes 
very rapidly to $\sim -75 \%$ at $\sim 20$ ps. 
With increasing excitation density $\wp$ reaches very large negative values, 
saturating at $\sim -90 \%$ for excitation densities larger 
than $20 W/cm^{2}$ (see inset of Fig. 2 b). 
The $\sigma ^+$-polarized excitation still creates a larger +1 spin population, 
which is reflected as a positive $\wp$ at t = 0. However, the scattering to 
\textbf{K} $\sim 0$ states of $-1$ spin becomes more efficient at longer 
times (t $\sim 20$ ps), resulting in a larger $-1$ spin population at \textbf{K} $\sim 0$ 
and therefore to a counter-polarized ($\sigma ^-$) emission.

In order to explain the negative values of polarization, one could argue that changing 
the excitation energy, going from $\delta > 0$ to $\delta < 0$, a resonant 
excitation condition with the light-hole excitons is met. 
However, this explanation can be disregarded because the excitation energies in 
our experiments are always above the light-hole exciton 
(at least 30 meV) and furthermore, a negative value of $\wp$ at t = 0 
would be obtained. The qualitative argument used to understand the spin dynamics 
for $\delta > 0$ does not apply anymore for the case of $\delta < 0$, since the seed 
for stimulation would still have +1 spin, resulting in positive values of $\wp$.

A detailed study of the polarization-resolved PL spectra clarifies the origin 
of the different spin dynamics for positive and negative detunings: 
a small energy splitting ($\Delta = E^{-} - E^{+}$, where $E^{+/-}$ denotes 
the $\sigma ^{+/-}$ emission energy) 
between the $\sigma ^+$ and the $\sigma ^-$ components of the 
PL is obtained at short delay times, as shown in Fig. 3. 
This splitting evidences that the +1 and $-1$ spin states are no longer 
degenerate in energy at \textbf{K} $= 0$.

The polarization-resolved PL spectra at 20 ps delay for 
$\delta$ = 10 meV are depicted in Fig. 3 (a): 
the $\sigma ^+$ emission (filled circles, solid line) occurs at lower 
energy than the $\sigma ^-$-polarized one (open circles, dashed line). 
The spin splitting, $\Delta$ (inset of Fig. 3 a) increases with 
excitation density, saturating at $\sim 1$ meV. On the contrary, in the case 
of $\delta < 0$ (Fig. 3 b), the $\sigma ^-$ component lies 
at lower energies, revealing that the $-1$ spin state is the lowest energy state 
at very short times. In this case $\Delta$ is negative and saturates 
at $\sim 0.5$ meV (inset of Fig. 3 b).

Let's describe in more detail the relaxation process of the non-resonantly 
created excitons towards \textbf{K} $= 0$ states, taking into account 
the $+1/-1$ spin splitting, and the fact that polariton pair-scattering is
spin selective.\cite {Savvidis00,Tartakovskii99,Renucci01,Martin01}
For $\delta > 0$, 
the large wave vector exciton-like polaritons from the LPB states are 
scattered to the cavity-like UPB \textbf{K} $= 0$ states. 
Most of the polaritons are transferred to the 
lowest energy spin level, i.e. +1 states, creating the seed for stimulation. 
This accumulation of +1 spin polaritons results in a large $\sigma ^+$-polarized 
stimulated emission and a considerably smaller $\sigma ^-$-polarized one: 
$\wp$ is positive and becomes very large at short 
times (see Fig. 2c). The +1 spin state is emptied very quickly through 
the $\sigma ^+$-polarized stimulated emission and therefore the 
polarization decreases to zero very rapidly after reaching the maximum. 
This balance of the populations, which is reinforced by the conventional 
spin relaxation processes, equalizes the intensities of both circularly 
polarized components of the PL and $\wp$ remains at zero.

For $\delta < 0$, the lowest energy level at \textbf{K} $\sim 0$ 
is now the $-1$ spin state. The accumulation of polaritons in those states 
results in a large $\sigma ^-$ emission. Therefore $\wp$ 
becomes negative and very large at short times. 
Similarly to the $\delta > 0$ case, now the $-1$ spin population is rapidly 
reduced and $\wp$ goes back to zero.
 
The physical origin of this energy splitting between the two spin states 
at \textbf{K} $\sim 0$ still needs to be clarified, but it is likely to 
account for the reversal of the circular degree of polarization of the PL with changing 
the exciton-cavity detuning. The splitting would be compatible with a 
decrease in the light-matter interaction strength for the majority polaritons (+1), 
which are initially created by the $\sigma ^+$-polarized excitation, 
as compared to that of the minority ($-1$) polaritons. 
This would imply that for negative (positive) detuning, the +1 states 
would lie above (below) the $-1$, rendering a $\Delta <0 (> 0)$ as borne out by our 
results. However, our experiments show that with increasing 
excitation density, an initial blue shift of 0.5 meV for 
both $\pm 1$ states, without any splitting, is 
followed by a red shift of the $-1$ polaritons, 
while the +1 remain at the same energy. Therefore, the coupling 
strength of the +1 polaritons does not decrease, invalidating the 
previous argument. The fact that the splitting increases with 
excitation power density indicates that it could originate 
from exciton-exciton interactions. An existing theory for bare excitons 
would qualitatively explain the splitting and 
the $\pm 1$ level ordering, as a result of exchange and vertex 
corrections to the 
self-energies,\cite{Joaquin96} but only for 
$\Delta < 0$. Further experiments are underway to 
understand this spin splitting and its \textbf{K} dependence.

In summary, we have investigated the relaxation and the spin dynamics of 
cavity polaritons after non-resonant pulsed excitation, 
in the non-linear emission regime. An increase of the excitation 
density leads to an exponential growth of the integrated emission intensity from the cavity-like states, 
providing an experimental evidence for the existence of final-state 
stimulated scattering. The spin dynamics presents novel phenomena, 
such as the existence of a maximum at a finite time and a sign 
reversal of the circular degree of polarization. This reversal is 
related with the sign of the splitting between the energies of the $\sigma ^+$- 
and $\sigma ^-$-polarized components of the PL. 
The spin of the lowest photon-like energy state changes
from +1 for $\Delta > 0$ to $-1$ for $\Delta < 0$.

\acknowledgments
This work has been partially supported by the EU (TMR-Ultrafast Quantum Optoelectronics Network), 
the Spanish DGICYT (PB96-0085) and the CAM (07N/0064/2001). 
We thank A. Kavokin for helpful discussions.

* Present address: Dept. of Physics and Astronomy, University of Southampton, SO17 1BJ Southampton, UK

\pagebreak

\begin{figure}
\caption{Contour plot of the PL at 10 ps measured in different points 
of the sample. Black = high intensity, light gray = low intensity. 
The dashed lines are guides to the eye. The arrows indicate the
detunings discussed in the text.
Inset: integrated intensity (Log scale) of the LPB emission at 20 ps as a 
function of the excitation power density for $\delta = -10$ meV. The arrow points the threshold for 
observing non-linear effects in the emission.}
\label{contour}
\end{figure}

\begin{figure}
\caption{(a) Time evolution of the circularly polarized PL of the UPB at $\delta = 10$ meV.
The filled circles/solid line (open circles/dashed line) 
denote the $\sigma ^+$ ($\sigma ^-$) emission. (b) Same as in (a) for the LPB at $\delta = -10$ meV.
Inset: maximum value of the polarization 
degree at 20 ps as a function of excitation density for $\delta= -10$ meV. 
The line is a guide to the eye. (c) Time evolution of the circular polarization degree of the PL emission 
of the UPB  at $\delta = 10$ meV.
(d) Same as in (c) for the LPB at $\delta = -10$ meV. 
The data are taken with an excitation density of $18 W/cm^{2}$}
\label{Polarization}
\end{figure}

\begin{figure}
\caption{(a) Polarization-resolved PL spectra at 20 ps for an excitation 
density of $18 W/cm^{2}$ at $\delta = 10$ meV. 
The filled circles/solid line (open circles/dashed line) 
denote the $\sigma ^+$ ($\sigma ^-$) emission. The lines are 
gaussian fits of the experimental data. Inset: Spin splitting 
($\Delta$) at 20 ps as a function of excitation density for $\delta$ = 10 meV. 
The line is a guide to the eye. (b) Same as in (a) for $\delta$ = -10 meV.}
\end{figure}
\label{Spectra}


\begin{references}
\bibitem{Weisbuch92}
{C. Weisbuch {\it et al.}}, Phys. Rev. Lett. {\bf 69}, 3314 (1992).
\bibitem{Houdre95}
{R. Houdre {\it et al.}}, Phys. Rev. B {\bf 52}, 7810 (1995).
\bibitem{Haroche89}
{S. Haroche, and D. Kleppner}, Physics Today {\bf 42}, 24 (1989).
\bibitem{Senellart99}
{P. Senellart, and J. Bloch}, Phys. Rev. Lett. {\bf 82}, 1233 (1999).
\bibitem{Stevenson00}
{R. M. Stevenson {\it et al.}}, Phys. Rev. Lett. {\bf 85}, 3680 (2000).
\bibitem{Dasbach00}
{G. Dasbach {\it et al.}}, Phys. Rev. B {\bf 62}, 13076 (2000).
\bibitem{Dang98}
{D. Le Si Dang {\it et al.}},Phys. Rev. Lett. {\bf 81}, 3920 (1998).
\bibitem{Bleuse98}
{J. Bleuse {\it et al.}}, J. Crys. Growth {\bf 184-185}, 750 (1998).
\bibitem{Boeuf00}
{F. Boeuf {\it et al.}}, Phys. Rev. B {\bf 62}, R2279 (2000).
\bibitem{Mueller00}
{M. Mueller, J. Bleuse and R. Andre}, Phys. Rev. B {\bf 62}, 16886 (2000).
\bibitem{Savvidis00}
{P. G. Savvidis {\it et al.}}, Phys. Rev. Lett. {\bf 84}, 1547 (2000).
\bibitem{Damen91}
{T. C. Damen {\it et al.}}, Phys. Rev. Lett. {\bf 67},  3432  (1991).
\bibitem{Vinattieri93}
{A. Vinattieri {\it et al.}}, Appl. Phys. Lett. {\bf 63}, 3164 (1993).
\bibitem{Ferreira94}
R. Ferreira, and G. Bastard, Solid State Electronics {\bf 37}, 851 (1994).
\bibitem{Maialle93}
M. Z. Maialle, E. A. de Andrada e Silva, and L. J. Sham, Phys. Rev. B {\bf 47}, 15576 (1993).
\bibitem{Tartakovskii99}
{A. I. Tartakovskii {\it et al.}}, Phys. Rev. B {\bf 60}, R11293 (1999).
\bibitem{Renucci01}
{P. Renucci {\it et al.}}, Springer Proceedings in Physics {\bf 87}. Edited by N. Miura and 
T. Ando. (Springer-Verlag, New York 2001), p. 653.
\bibitem{Martin01}
M. D. Mart\'in, L. Vi\~na, R. Ruf, and E. Mendez, Solid State Comm. {\bf 117}, 267 (2001).
\bibitem{Savvidis02}
{P. G. Savvidis {\it et al.}}, Phys. Rev. B {\bf 65}, 73309 (2002).
\bibitem{Aichmayrxx}
G. Aichmayr, M. D. Mart\'in, L. Vi\~na, and R. Andr\'e, unpublished.
\bibitem{Baumberg00}
{J. J. Baumberg {\it et al.}}, Phys. Rev. B {\bf 62}, R16247 (2000).
\bibitem{Joaquin96}
{J. Fern\'andez-Rossier {\it et al.}}, Phys. Rev. B {\bf 54},  11582  (1996).


\end{references}
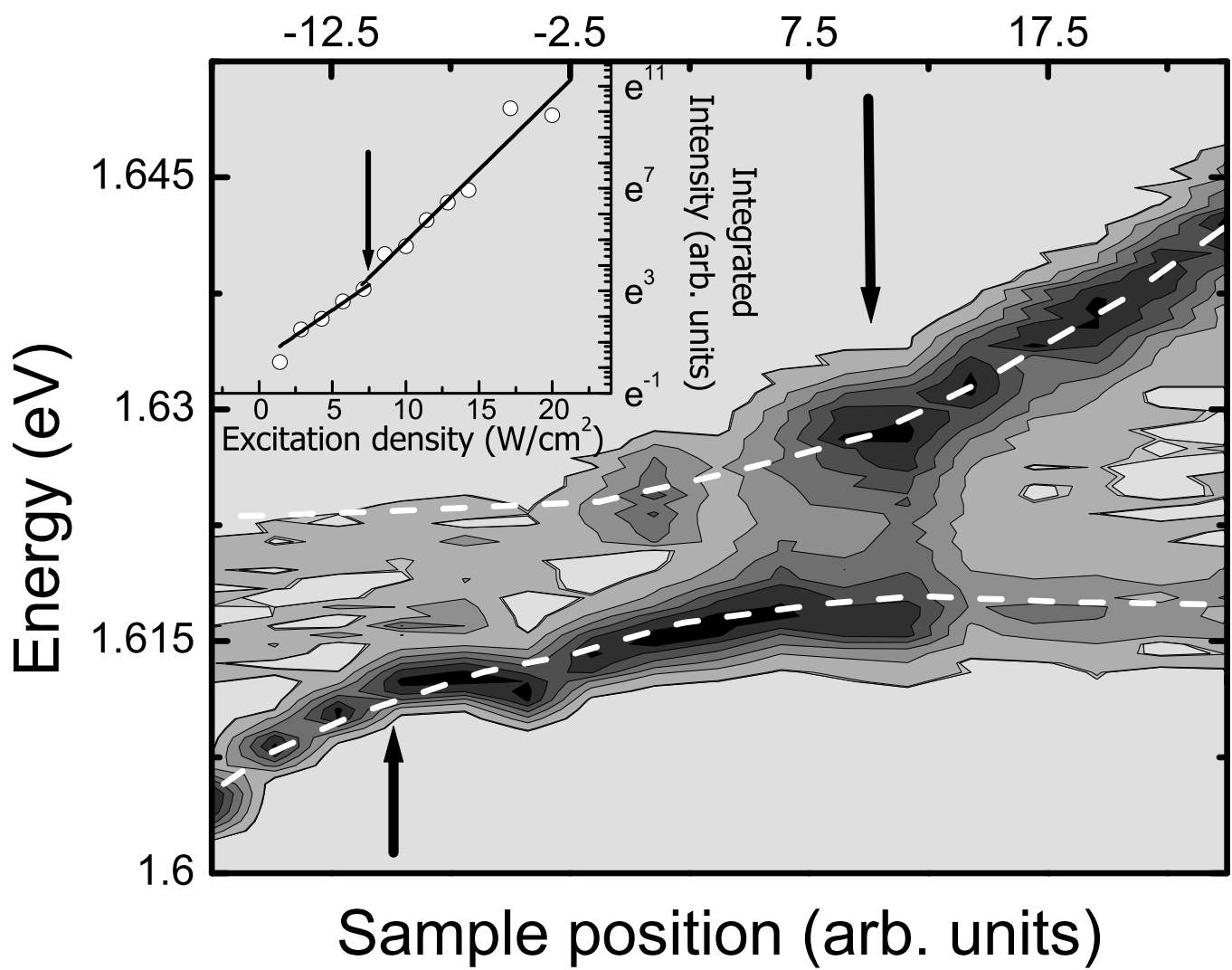
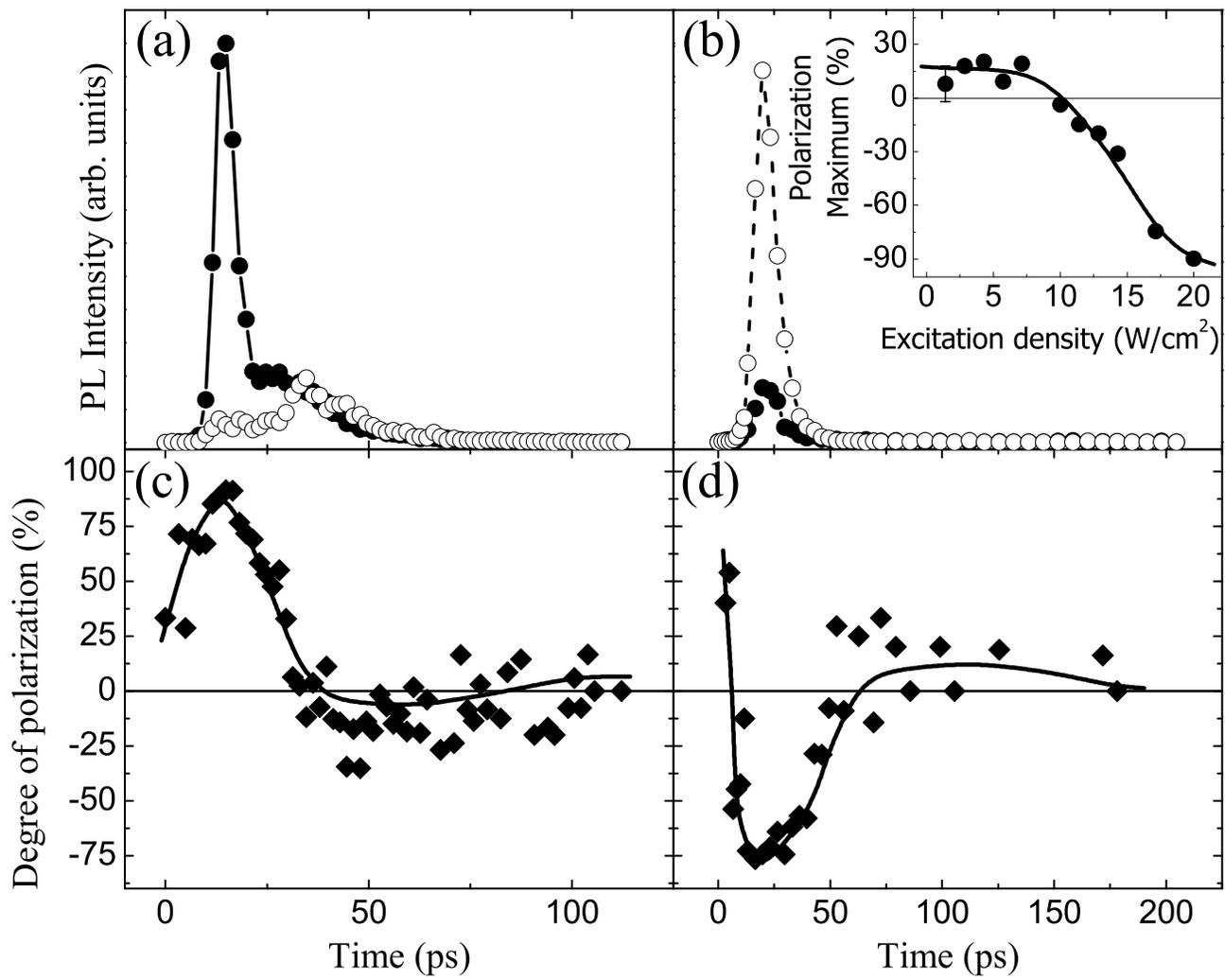
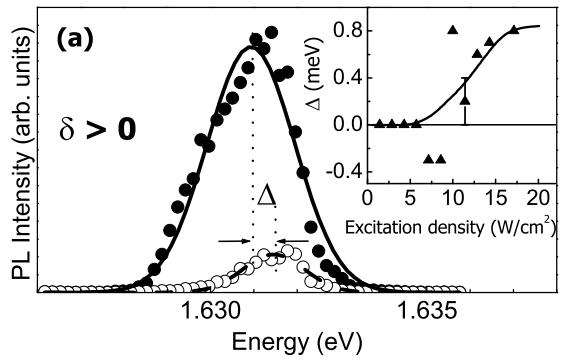
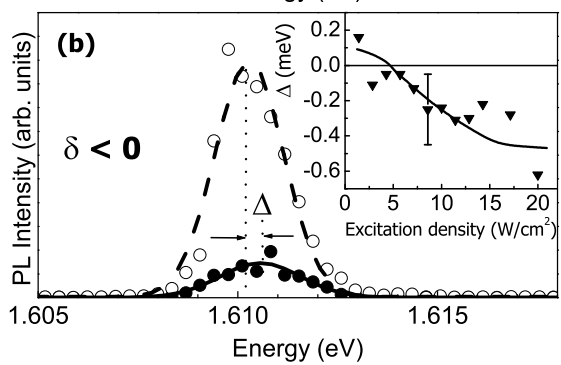
\end{document}